\documentclass{an}
\usepackage{graphicx}
\usepackage{times}
\usepackage{fancyhdr}
\begin{document}

\title{Soft gamma repeaters activity in time}

\author{S.B. Popov\inst{1}\fnmsep\thanks{Corresponding author:
  \email{polar@sai.msu.ru}}
}
\titlerunning{Soft gamma repeaters activity in time}
\authorrunning{S.B. Popov}
\institute{Sternberg Astronomical Institute, Universitetski pr. 13,
Moscow 119991, Russia}

\received{}
\accepted{}
\publonline{later}


\abstract
{In this short note I discuss the hypothesis that bursting activity of
magnetars evolves in time analogously
to the glitching activity of normal radio pulsars
(i.e. sources are more active at smaller ages), and that the increase of the
burst rate follows one
of the laws established for glitching radio pulsars.
If the activity of soft gamma repeaters decreases in time in the way 
similar to the evolution of core-quake glitches ($\propto t^{5/2}$),
then it is more
probable to find the youngest soft gamma repeaters, but the energy of giant
flares from these sources should be smaller than observed
$10^{44}$~--$10^{46}$~ergs as the total energy stored in a magnetar's
magnetic field is not enough to support thousands of bursts similar to the
prototype 5 March 1979 flare.}


\maketitle


\section{Introduction}

 Soft gamma repeaters (SGRs)
are known to demonstrate three main types of activity: weak,
intermediate, and giant flares (see a recent review in Woods \& Thompson
2006). \footnote{Here often I do not discuss separately, 
in the context of the proposed hypothesis,
hyperflares (HFs) like the one from SGR 1806-20
observed on 2004 Dec 27. Such events are just included into the class of giant flares.}
More energetic events are less frequent than weak bursts.
The distribution of SGRs burst energy has a power-law shape, $dN/dE \propto
E^{-\alpha}, \alpha\sim 5/3$. This distribution is similar to the
Gutenberg-Richter law established for terrestrial quakes (Cheng et al. 1996; G\"o\u{g}\"u\c{s} et al. 1999).
A lot is known about properties of bursts of different kinds, but the
evolution of the rate of bursts during lifetime of SGRs in not known due to
poor statistics and relatively short period of exploration of these sources.
 
Here I discuss a possible link between SGRs bursts and period glitches,
especially
glitches driven by starquakes. Such connection was suggested long
ago.
Also the possibility that starquakes can power cosmic gamma-ray bursts was
discussed by
Blaes et al. (1989) (see references to earlier papers therein).
 However, to the author's knowledge, the possible similarity between the
evolution of the bursts rate during a SGR lifetime  and
the evolution of the rate of glitches derived from radio
pulsar observations has never been discussed in literature.
The reason lies in the accepted difference between radio pulsar and magnetar
glitches. In the case of
 radio pulsars,  two types of glitches are distinguished: Vela-like and
Crab-like. They are thought to be due to either starquakes or superfluid
vortex unpinning. Glitches in magnetars are believed to 
be connected with strong
magnetic fields of these stars. Nevertheless, I think that a discussion of
possible similarities between  the evolution of glitch rates in different
kinds of neutron stars (NSs) is worth while, especially in connection with
giant flares.

During $\sim 25$~--~$30$ years of observation four strong flares
have been detected from four known SGRs (Mazets et al.
2004)\footnote{Often only two bursts -- 05 March 1979 from SGR 0526-66 and
27 August
1998 from SGR 1900+14 -- are called giant bursts. Many authors do not include the flare from SGR
1627-41 on 18 June 1998 into the list of giant flares, as for this event a
pulsating ``tail'' was not observed. The flare of SGR 1806-20 is also sometimes
classified as an event of a different type -- a {\it hyperflare} - HF.}. 
Despite their smaller rate, due to large energy release
these bursts dominate among other types of bursting activities of
SGRs. The rate of such flares was usually assumed to be $\sim
0.02-0.01$ per year per source (Woods \& Thompson 2006) based on
just two giant flares (GFs) (5 March 1979 and 28 August 1998). However, with
inclusion of the 27 December 2004 flare and with inclusion of the 18
June 1998 burst of SGR 1627-41 
the rate becomes higher. Still, below I shall use a
conservative estimate $0.02$ GFs per year per source. Data on
possible extragalactic SGR flares (Ofek 2007) is only marginally 
compatible with this rate, and I briefly discuss this problem below.

 Previously several authors (Nakar et al. 2006; Popov \& Stern 2006;
Lazzati, Ghirlanda \& Ghisellini 2005; Ofek 2007) addressed the problem of
extragalactic SGR flares. They used different archive data (BATSE, IPN) to
search for examples of extragalactic GFs and HFs, and to put limits on the
fraction of SGR flares among short GRBs (see Nakar 2007 for a review on
short GRBs and for some discussion of the fraction of SGR flares among
them). Estimates show that GFs can be observed by most of the detectors
(from BATSE to SWIFT) up to distances of few Mpc (Palmer et al. 2005). I.e.,
even SWIFT cannot observe GFs from most of the Virgo cluster -- only HFs can
be detected from such distance (Hurley et al. 2005). \footnote{Details about
SWIFT sensitivity can be found in (Band 2006). SWIFT appears to be more
sensitive than BATSE to long bursts, but not for short hard events. The
advantage of SWIFT is in better localization of short events. BAT onboard
SWIFT is less sensitive than BATSE to energies $>100$ keV. HFs are
relatively hard events: spectra of the HF 27 December 2004 can be described
as a blackbody with T~$\sim$~200 keV. So, in general, SWIFT is not much
better than BATSE for detection of bursts like the only known HF. Other
experiments, like BeppoSAX and HETE-II, also are not more sensitive than
BATSE to short hard bursts. However, as noted by Band (2006), an analysis of
short-duration triggers by SWIFT which do not result in successful image
triggers can result in discovery of more short flares.
 Up to the limiting distance for SWIFT detection the extrapolation of the galactic rate of GFs gives an
estimate of few  events in a year, so SWIFT with its sky coverage of about 1.4
steradians can detect approximately one GF per year. 
HFs can be detected from larger
distances: 30-40 Mpc by BATSE, and up to 70 Mpc -- by SWIFT (Hurley et al.
2005). Scaling the galactic rate of HFs -- one in 30 years -- Hurley et al.
(2005) gives the rate about 50 per year for SWIFT.
However, any rescaling of the galactic rate of HFs
is extremely uncertain as only one event has been observed. 
Results presented in (Popov \& Stern 2006; Lazzati
et al. 2005; Ofek 2007) show that the rate of extragalactic
HFs is much lower than that  expected from a
 naive  scaling (based on starformation rate or supernova rate, etc.) given
one observed event from SGR 1806-20 in our Galaxy. Upper limits from
non-detection of HFs from the Virgo cluster by BATSE provide an estimate
$\sim 10^{-3}$ yr$^{-1}$ per Milky Way-like galaxy (Popov \& Stern 2006).
This number after rescaling again gives an estimate of about one detection
per year for SWIFT. Still, larger statistics of HFs detection is necessary.
Recent observations (Frederiks et al. 2007; Ofek et al. 2006) have already
provided a candidate for a HF in a close-by galaxy with high starformation
rate (such stellar systems are the most prominent hosts of GFs and HFs).
Hopefully, more events can be detected by SWIFT during its lifetime.}

Usualy authors apply the value of GFs rate $\sim 0.02-0.01$ 
per year per source to the whole life
period of SGRs, i.e. the rate of GFs is assumed to be constant in
time. In this short note I discuss an alternative 
assumption. NS activity of different kinds usually decreases with
time, and the GF rate should not be an exception. Possibly the
closest analogue is the evolution of glitch (especially,
starquake) rate.


\section{Possible evolution of the giant flare
rate during SGRs lifetime}

 Bursting activity of magnetars is known to be associated with glitches.
Such an association naturally arose already in the early papers by
Thompson \& Duncan (1996).
For SGRs themselves, period determinations are not  precise enough to allow
glitch detections\footnote{Also glitches can be masked by a transient period
increase as the one observed for SGR 1900+14, see discussion, for example,
in Thompson, Lyutikov \& Kulkarni (2002).},
but for anomalous X-ray pulsars (AXPs) such correlation
between bursts and glitches has been found.  
For example, the source  1E2259+586
showed a glitch coincident with a burst
(Kaspi et al. 2003; Woods et al. 2006).

 Despite the possibility that
glitches of SGRs and AXPs can be different from normal radio pulsar glitches
due to the role of superstrong magnetic fields,
here I want to propose the following {\it hypothesis}: 
bursting activity of magnetars
(mainly the GF rate)
evolves with time similarly to the glitching activity of normal radio pulsars.

 Glitches of radio pulsars are well known since 1969
(Radhakrishnan \& Manchester 1969; Reichley \& Downs 1969).
Relatively recent data on statistical properties of glitches can
be found in Lyne, Shemar \& Graham Smith  (2000). The most
up-to-date information is available on the Web in the ATNF
catalogue\footnote{http://wwwatnf.atnf.csiro.au/research/pulsar/psrcat/}.

Long-period pulsars also show glitches. 
The well-known long period glitching pulsar is PSR B0525+21. It has
$p=3.75$~s, $\dot p \approx4 \, 10^{-14}$, $B=1.24 \, 10^{13}$~G,
age $\sim 1.5 \, 10^6$~yrs. Recently, the glitch was reported
(Janssen \& Stappers 2006) from PSR J1814-1744 with $p\sim 4$~s,
$B\sim 5.5 \, 10^{13}$~G, $\dot p \sim 7.4\, 10^{-13}$, and age
$\sim 8.5\, 10^4$~yrs. So, probably, magnetars which have somehow similar
periods can also suffer normal (i.e. not magnetically driven)
glitches. It is interesting to note, that as there are claims that
SGRs are born from the most massive stars which still can produce
NSs (Figer et al. 2005; Gaensler et al. 2005),
then for them it is possible to apply the consideration (see, for example,
Alpar \& Ho 1983)
that more massive NSs can produce more rare, strong glitches.
This becomes posible due to a solid core formation
if a very stiff equation of state is realized in nature.
Formation of a solid core necessary to produce very large glitches can be an
interesting option distinguishing between SGRs and ``normal'' NSs.

 Recently, a glitch was proposed to interpret the behavior of 
one of the {\it Magnificent Seven}
-- radioquiet close-by cooling NSs -- RX J0720-3125 (van Kerkwijk et al.
2007). It is a NS with the spin period 8.4 s. If the glitch of RX J0720-3125 
is due to unpinning, then using standard formulae
(Alpar \& Baykal 1994) it is possible to estimate the reccurence time of
glitches in this source.
Surprisingly, the time is about 10 years. So, it is quite probable to
observe one since 1997.\footnote{In this source glitches, probably, generate
Tkachenko waves (Tkachenko 1966, Ruderman 1970), 
which in their turn lead to precession.} 
So, even for long-period NSs which do not show radio pulsar behaviour  
glitches can be a normal phenomena. 

Two main models of glitches are discussed:
starquakes (Ruderman 1969) and vortex unpinning (Anderson \& Itoh 1975).
The first of these two models can be  especially good analogue of
magnetar glitches.
Let us discuss the evolution of glitches rate according to these two well-known
models.

 For the model of unpinning of superfluid vortices
Alpar \& Baykal (1994)
demonstrated arguments in favour of the following glitch
property: for all NSs the ratio $\delta \Omega / \Omega$ is
nearly the same. Where $\Omega$ is a spin frequency, and $\delta \Omega$
is the parameter which 
determines the time interval between glitches according to the following
simple relation:

{\setlength{\mathindent}{0pt}
\begin{equation}
t_{\mathrm{g}}=\frac{\delta \Omega}{\dot \Omega}.
\label{eq1}
\end{equation}}

Parameter $\delta \Omega$ is the critical value of the difference between
the rotation frequencies of normal matter and the superfluid at a boundary
layer.
$\delta \Omega$ itself can be estimated as (Alpar \& Baykal 1994; Cheng \&
Chi 1996):

{\setlength{\mathindent}{0pt}
\begin{equation}
\delta \Omega = \Delta \Omega \frac{I_0}{I_{\mathrm{p}}},
\end{equation}}
here $I_{\mathrm{p}}$ is the effective moment of inertia of the region of
a pinning layer.

Combining these two formulae we obtain the relation for 
the time between glitches:

{\setlength{\mathindent}{0pt}
\begin{equation}
t_{\mathrm{g}} =
\frac{2\, I_0}{I_{\mathrm{p}}} \frac{\Delta \Omega}{\Omega} t \propto t,
\label{eq2}
\end{equation}}
where $t$ is the age of a pulsar, 
$t=\Omega/2\, \dot \Omega$. To obtain the
proportionality
$t_{\mathrm{g}}\propto t$
it is assumed that $\Delta \Omega / \Omega$ is constant during  a NS lifetime.

If for the age $t_0=1000$~--~$2000$ yrs a SGR has GF rate equal to one
burst in $\sim $~25~--~50 years, then for the age of about 100 yrs
the rate is about one burst in 1-5 years. It is unclear whether a NS
can produce bursts at earlier time since only at the age 50-100
yrs it becomes isothermal (Prakash et al. 2001).

However, eq.~\ref{eq1} is valid for the superfluid vortex unpinning model
(Alpar \& Baykal 1994), which is unlikely to be valid for magnetars.
For the core-quake model, the formula for $t_{\mathrm{g}}$ is
different (Alpar \& Ho 1983; Alpar \& Baykal 1994):

{\setlength{\mathindent}{0pt}\begin{equation}
t_{\mathrm{g}} = \frac{2(A+B)\phi \Delta \Omega / \Omega}
{I_0\Omega \dot \Omega}.
\label{eq3}
\end{equation}}

This scenario gives a very quick rise of the glitch rate towards the pulsar's
youth:

{\setlength{\mathindent}{0pt}\begin{equation}
t_{\mathrm{g}} \propto t^{5/2},
\label{eq4}
\end{equation}}
if $A$, $B$, $\phi$,
and $\Delta \Omega$ are constant. If $\Delta \Omega$ decreases with
time, then the dependence is slightly less steep. For example, for  $\Delta
\Omega / \Omega=$~const we obtain $t_{\mathrm{g}} \propto t^2$.
Therefore, if the rate of GFs follows the same law (for example, glitches
can trigger GFs), then young SGRs produce GFs much more often that the
observed 1000-year old sources. Magnetic field decay can lead to a 
dependence of the glitching period on time even steeper than that in eq.~\ref{eq4}.

In section 3 I briefly discuss some consequencies of this hypothesis.

\subsection{$\Delta \Omega/ \Omega$ estimate}

Using eqs.~\ref{eq2} or \ref{eq3} it is possible to obtain a naive estimate 
of the value $\Delta \Omega/ \Omega$ after a GF. 
Time intervals between GFs are about 50 years. 
In the case of eq.~\ref{eq2} assuming $I_{\mathrm{p}}=10^{-2} I_0$
(Alpar \& Baykal 1994; Cheng \& Chi 1996) we
obtain $\Delta \Omega/ \Omega \sim 10^{-4}$.
This is a very large value. 

In the case of eq.~\ref{eq3} for $A=10^{52}$ erg, $\phi=10^{-3}$,
$B=10^{48}$~erg, $\Omega \sim 1$~rad~s$^{-1}$, $\dot \Omega \sim
10^{-11}$~rad~s$^{-2}$, and $I_0\sim 10^{45}$~g~cm$^2$ we obtain
$\Delta \Omega/ \Omega \sim 10^{-6}$. Characteristics of the
glitch of 1E 2259+586 are consistent with this estimate by Kaspi
et al. (2003). Of course, the use of parameters similar to radio
pulsars is an oversimplification.  For starquakes, in the case of
magnetars $\Delta \Omega/ \Omega$ hardly can  be that high as it should be
limited by oblateness: $\epsilon \sim \Delta \Omega/ \Omega $ and
$\epsilon \sim a/g \sim {\mathrm {few}} \, 10^{-9} \, \Omega^2$,
where $g$ is the gravitational acceleration and $a$ is the
centrifugal one.  Smaller $\Delta \Omega/ \Omega$ correspond to smaller
$t_\mathrm{g}$. For realistic $\Delta \Omega/ \Omega$ and other parameters
as specified above glitches should be more frequent. This can mean that not
every glitch lead to a GF.

\section{Discussion}

\subsection{Detectability of SGRs of different age}

 Known SGRs have spin-down ages 
of about 1-2 thousand years (Woods \& Thompson
2006).
It has been claimed that observations of younger SGRs are less probable due to
a smaller number of such objects.
However, if our hypothesis is correct it is not so!
If the bursting activity is scaled in linear proportion to a source's age, then
there is an equal probalitity to find a SGR of any age. If $t_{\mathrm{g}}$
inreases with time faster than it is described in eq.~\ref{eq2}
then the probability to
detect a younger SGR is even higher, despite that the number of such sources
is small.  This means that 
the probability of detection is biased towards younger objects.

The fact that we do not detect younger SGRs in our Galaxy (and in
near-by galaxies like Magellanic clouds) tells us, that there are
no younger sources of this type in our close proximity,
 as they should burst more often than the four known sources, and in our
vicinity we cannot miss even weak SGR flares. 
In the case of eq.~\ref{eq4} it is especially obvious  as the increase of the burst
rate for young sources is extremely strong. Or, one can turn 
the argument around, the absence of other known close-by SGRs can rule out
significant decrease of flare rate in time, if the rate of
formation of these sources in our neighborhood is larger than one
in several hundred years (on the formation rate of magnetars see the recent
paper by Gill \& Heyl 2007). 

 If activity of SGRs decays with time, then bounds obtained in
(Popov \& Stern 2006; Lazzati et al. 2005; Ofek 2007)
become stronger, as the total number of bursts of any
kind is larger than the one derived by naive use of the observed
(from the four sources) rate in the form of a constant value valid for the
whole lifetime of a magnetar, unless GFs and HFs from younger SGRs are somehow
weaker.

\subsection{Powerful flares of young SGRs and energy crisis}

 There are suggestions (see, for example, Hurley et al. 2005) that younger
magnetars can produce flares even more powerful than the one
observed on December 27, 2004. This suggestion is based on the
fact that young SGRs must have shorter spin periods and higher
magnetic fields. If so, then we can expect to find such
young SGRs in starforming galaxies (Popov 2005; Popov \& Stern
2006), and possibly two candidates have recently been observed (Frederiks et
al. 2006; Ofek et al. 2006; Golenetskii et al. 2007). 
However, if bursts of young magnetars appear much more
often, then it is necessary to estimate whether there is enough energy
resources for the increased rate of more powerful flares. It is
important to note here, that as weak bursts seem to cluster around
more energetic ones we do not expect to observe just an increase
of weak bursts activity without corresponding increase in the
number of GFs, and, vice versa, an increase of only GFs rate is
unexpected, too.

Suppose that the formula~\ref{eq2} is valid. Then with the present-day
rate of GFs of about one in 50 yrs per source, we can expect, that a
NS in the time interval $100<t<2000$ can produce about 100 GFs.
The total energy stored in the magnetic field can be estimated as:

{\setlength{\mathindent}{0pt}
\begin{equation}
E_{\mathrm{tot}} = 2\, 10^{47} B_{15}^2 \, {\mathrm{erg}}.
\label{eq5}
\end{equation}}
 So, there is enough energy to
produce 100 GFs with energy of $\sim 10^{45}$~erg. However, there is
not enough energy to produce significantly more powerful flares in
a SGR youth. To produce a comparable amount of more
energetic flares it is necessary to have initial field of about
$10^{16}$~G (Stella et al. 2005). This is possible if young magnetars have
much higher magnetic fields than the observed ones (see, Pons \& Geppert
2007).

As supernova rates in galaxies with extreme starformation can be
up to 100 times higher than that in the Milky Way,  in such systems
we can expect several SGRs with ages of about 100 years. 
Then the rate of GFs from such ``supernova factories'' can be
one in few months. Non-detection of such bursts or at least small
number of candidates coincident with expectations of random
projections (Popov 2005) can be an additional argument against
frequent HFs from young SGRs.

In the case of viability of the eq.~\ref{eq4} the situation is different. We
can expect $> 10^3$ GFs during the interval $100<t<2000$~years. In
this case, even for constant energies of flares we face an energy
crisis. For an increase of a flare energy towards smaller ages the
crisis becomes even more dramatic. The energy of a glitch itself
is too low to contribute significantly to a GF:

{\setlength{\mathindent}{0pt}
\begin{equation}
E = I_0 \Omega \Delta \Omega = 10^{40} \Omega_1^2
\left(\frac{\Delta \Omega}{\Omega}\right)_{-5} \, {\mathrm{erg}},
\end{equation}}
here $\left(\frac{\Delta \Omega}{\Omega}\right)_{-5}=
(\Delta \Omega / \Omega)/10^{-5}$.
In the case of magnetars $\Delta \Omega/ \Omega$ after a starquake
can not be that big as it is limited by oblateness that is small
for SGRs spin periods, $ \Delta \epsilon \le \epsilon \sim
{\mathrm{few}} \,10^{-9} \Omega^2.$ However, estimates of
oblateness for RX J0720.4-3125 (one of the so-called {\it
Magnificent Seven}) is $\sim 4\, 10^{-8}$ (Haberl 2007).
Even higher values of deformation can be reached due to strong toroidal
field $B_\mathrm{t}$.
Stella et al. (2005) showed
that the value of ellipticity can be as high as 
$\sim 6.4 \, 10^{-4} B_{\mathrm{t, 16.3}}^2$. However, this ellipticity
can be unrelated to glitches.


Thermal energy of a NS after few hundred years is not enough to support a
magnetar activity.
The total rotational energy is also unsufficient to
support significant number of GFs when a magnetar slowed down:
$E_{\mathrm{rot}} \approx 10^{45} I_{45} \Omega^2$.

So, if the rate of GFs follows the same decay with time as the
rate of radio pulsar glitches  in the starquake model, an
avegare energy of GFs of young SGRs should be smaller in
comparison with older objects in order not to face the energy
crisis. These considereations can be important for estimates of
detectability of extragalactic GFs (see discussion of this subject
in Duncan 2001; Popov \& Stern 2006; Lazzati et al. 2005, Crider 2006; Ofek
2007 and references therein).

Let the present day typical energy in a giant flare be $E_0$, and
the interval between such flares $k_0$. Let them vary as: $k_0
(t/t_0)^a $ and $E_0 (t/t_0)^{b}$. Naively one can expect $a>0$
and $b<0$. Then the total emitted energy is:

{\setlength{\mathindent}{0pt}
\begin{equation}
E_{\mathrm{em}}= \int_{t_1}^{t_2} \frac{E_0 (t/t_0)^{b}}{k_0
(t/t_0)^a} dt.
\label{eq6}
\end{equation}}

For $b=0$ (i.e. constant flare energy) the energy crisis appears
for $a\sim 2.5$ for $k_0=50$~yrs and $E_0=3\, 10^{44}$~erg. For
$b<0$ the crisis appears even for smaller $a$.

It is important to note, that the
evolution of the giant bursts rate cannot solve the problem
of the deficit of events in the direction of the Virgo cluster.
The contradiction between galactic and extragalactic rates
discussed in (Popov \& Stern 2006) is based on the observed
rate of GF in our Galaxy and its rescaling with the supernova rate.
Smaller number of energetic 
HFs from very young magnetars cannot change
the total rate significantly.


\subsection{SGRs -- ultrahigh energy phenomena -- starformation galaxies}

 Eichler (2005) recently proposed (see also Singh, Ma \& Arons 2004) that GFs can be an important source of
ultrahigh energy particles. In particular, the author discussed a
possibility of ultrahigh energy cosmic rays (UHECR) generation
during these bursts. Increased bursting activity of young SGRs
 can be responsible
for even higher contribution  to UHECR production. In this respect
it is interesting to remember the result by Giller, Michalak \&
Smialkowski (2003)
These authors claim that there are correlations between UHECR and
ultra-luminous IR galaxies (ULIRGs). In particular, these authors
discussed the galaxies Arp 299 and NGC 3256. The same objects were
also suggested by us (Popov 2005; Popov \& Stern 2006) as good
candidates to host multiple young SGRs. 
I would like to underline a possible role of very young magnetars,
mainly populating starforming regions, in a hypothetical UHECR
production, and their strong link with galaxies with high
starformation rate, especially with ULIRGs. If the rate of energy
release is larger for young SGRs then these sources can contribute
to the production of energetic particle in starforming galaxies on the level
higher than  estimated from the known galactic population of SGRs,
i.e. naive scaling based just on the number of SGRs is not valid.
Still, the total contribution of magnetars into the UHECR production is anyway 
limited by the total energy stored in their magnetic fields. 



\section{Conclusions}

 I propose and discuss
the hypothesis that the bursting activity of SGRs decreases
with time similar to the glitching activity of normal radio pulsars.
This assumption leads to several interesting consequences. Namely,

\begin{itemize}
\item For the same burst energy
the probability to detect a SGR at least does not diminish with
decreasing  SGR age. Even it can be more probable to
detect young SGRs due to their more frequent bursts. For example,
the galactic (in general, Local group) population of SGRs should
already include all the youngest objects of this type.
\item If the bursting activity of magnetars
rapidly decreases with time, but a typical GF energy is constant
(or decreases too), then we face an energy crisis: there is not enough energy
in the magnetic field to support thousands of 
flares similar to, say, the 5 March
1979 burst.
\item Young SGRs should burst more often, but their giant flares
should be less powerful than the known ones.
\end{itemize}

Absence of many identified extragalactic giant and hyper flares by itself
can rule out joint decrease of the flare rate and flare energy
with time.

\section*{Acknowledgments}

I want to thank Drs. V.A. Belokurov, M. Lyutikov,
M.E. Prokhorov and B.E. Stern
for comments.
The work was supported by the RFBR grant 06-02-16025
 and by INTAS.


\end{document}